\documentclass[12pt]{article}
\usepackage{epsfig,amsfonts,amsmath,array,mathrsfs}
\newcommand{\pbs}[1]{\let\temp=\\#1\let\\=\temp}
\renewcommand{\theequation}{\thesection.\arabic{equation}}
\def\be{\begin{equation}}\def\ee{\end{equation}}
\def\cvp{\raise 2pt\hbox{,}}

 \def\Tr{\mathop{\rm Tr}\nolimits}

\def\d{{\rm d}}
\def\nn{{\cal N}}

\def\Nf{N_{\mathrm f}}
 \def\uN{{\rm U}(N)} 
  
\def\La{\Lambda}

\def\wt{W_{\text{tree}}}

\def\q{\mathsf q}
\def\b{\boldsymbol{b}}
\def\g{\boldsymbol{g}}
\def\x{\boldsymbol{x}} 

\def\u{\text{U}(1)}
 
\def\ranglecl{\rangle_{\text{cl}}}

\def\plb#1#2#3{{\it Phys.\ Lett.\ }{\bf B #1} (#2) #3}
\def\npb#1#2#3{{\it Nucl.\ Phys.\ }{\bf B #1} (#2) #3}

\def\jhep#1#2#3{{\it JHEP\ }{\bf #1} (#2) #3}

\def\atmp#1#2#3{{\it Adv.\ Theor.\ Math.\ Phys.\ }{\bf #1} (#2) #3}

\begin{document}
%
%
\pagestyle{empty}
{\parskip 0in

\hfill LPTENS-07/48

\hfill arXiv:0710.2978 [hep-th]}

\vfill
\begin{center}
{\LARGE Consistency conditions in the chiral ring}

\medskip

{\LARGE of super Yang-Mills theories}

\vspace{0.4in}

Frank \textsc{Ferrari} and Vincent \textsc{Wens}
\\
\medskip
{\it Service de Physique Th\'eorique et Math\'ematique\\
Universit\'e Libre de Bruxelles and International Solvay Institutes\\
Campus de la Plaine, CP 231, B-1050 Bruxelles, Belgique
}\\
\smallskip
{\tt frank.ferrari@ulb.ac.be, vwens@ulb.ac.be}
\end{center}
\vfill\noindent

Starting from the generalized Konishi anomaly equations at the
non-perturbative level, we demonstrate that the algebraic consistency
of the quantum chiral ring of the $\nn=1$ super Yang-Mills theory with
gauge group $\uN$, one adjoint chiral superfield $X$ and $\Nf\leq 2N$
flavours of quarks implies that the periods of the meromorphic one-form
$\Tr\frac{\d z}{z-X}$ must be quantized. This shows in particular that
identities in the open string description of the theory, that follow
from the fact that gauge invariant observables are expressed in terms
of gauge variant building blocks, are mapped onto non-trivial
dynamical equations in the closed string description.

\vfill

\medskip
%
\begin{flushleft}
\today
\end{flushleft}
\newpage\pagestyle{plain}
\baselineskip 16pt
\setcounter{footnote}{0}

\section{Introduction}
\setcounter{equation}{0}

The fact that any four dimensional gauge theory has two seemingly
unrelated formulations, one in terms of open strings, which is
equivalent to the standard field theoretic Yang-Mills description and
the other in terms of closed strings, which thus contains quantum
gravity, is an extremely deep and fascinating property. Following
\cite{AdS}, many successfull examples of this duality have been
studied over the last decade. Yet many questions, both technical and
conceptual, remain unsolved.

A fundamental conceptual issue is to understand how the basic
ingredients in one formulation are encoded in the other formulation
and vice-versa. For example, how does the closed string gravity theory
know about the Yang-Mills equations of motion? In the closed string
description, we do not see the gauge group, for only gauge invariant
quantities can be constructed. This is of course not an inconsistency,
since the gauge symmetry is really a redundancy in the description of
the theory and not a physical symmetry. However, how then can we
understand charge quantization \`a la Dirac, which is usually derived
from gauge invariance, in the closed string set-up? A directly related
question, which will be at the basis of the present work, is the
following. In the open string framework, gauge invariant observables
are built in terms of fields that transform non-trivially under the
gauge group, and this has some non-trivial mathematical consequences.
For example, imagine that the gauge group is $\uN$ and that the theory
contains an adjoint field $X$. The gauge invariant operators built
from $X$ are obtained by considering traces
\be\label{ukdef} u_{k} = \Tr X^{k}\ee
or product of traces. The fact that $X$ is a $N\times N$ matrix
implies that there exists homogeneous polynomials $P_{p}$ of degree
$N+p$, if the degree of homogeneity of $X$ is one, such that
\be\label{defPp} u_{N+p} = P_{p}(u_{1},\ldots,u_{N})\, ,\quad p\geq
1\, .\ee
Thus only $u_{1},\ldots,u_{N}$ are independent. But how does the
closed string theory know about \eqref{defPp}, while the matrix $X$
does not exist in the closed string framework? In a sense we are
asking how to build the open strings starting from the closed strings,
which is a notoriously difficult question.

An extremely interesting incarnation of the open/closed string duality
is obtained when one focus on the chiral sector of $\nn=1$
supersymmetric gauge theories. The closed string set-up involves a
geometric transition \cite{GV} and is equivalent to the Dijkgraaf-Vafa
matrix model description \cite{DV}. On the other hand, the model has
been solved recently starting from the usual field theoretic
description \cite{mic1,mic2,mic3}, using Nekrasov's instanton
technology \cite{nekrasov}. The theory is essentially reduced to a
statistical model of colored partitions which, remarkably, yields
gauge theory correlators that coincide with the matrix model
predictions \cite{mic2,mic3}. The open/closed string duality is thus
fully understood in this case. Our aim in the present paper, which is
a continuation of \cite{ferchiral}, is to address some of the above
conceptual questions in this well-controlled framework. Our main
result will be to show that identities like \eqref{defPp} are
equivalent to dynamical equations of motion in the closed string
description.

The plan of the paper is as follows. In Section 2, we introduce some
basic ideas on a very simple example and present the model we are
studying, the $\nn=1$ super Yang-Mills theory with gauge group $\uN$,
one adjoint chiral superfield and $\Nf\leq 2N$ flavours of quarks. We
also state the chiral ring consistency theorem \cite{ferchiral}. This
is our main result and the proof of the theorem is given in Section 3.
Finally in Section 4 we summarize our findings and conclude.

\section{Preliminaries}
\setcounter{equation}{0}
\subsection{A simple example: the classical limit}
\label{classexSec}

We can immediately give the flavour of the arguments that we are going
to use by looking at the classical limit. We consider the $\uN$ super
Yang-Mills theory with one adjoint chiral superfield $X$ and
tree-level superpotential $\Tr W(X)$ such that
\be\label{wtdef}W'(z) = \sum_{k=0}^{d}g_{k}z^{k} =
g_{d}\prod_{i=1}^{d}(z-w_{i})\, .\ee
The equations of motion in the open string description are thus
\be\label{openclem} W'(X) = 0\, .\ee
The most general solution is labeled by the positive integers $N_{i}$,
with
\be\label{Nisum}\sum_{i=1}^{d}N_{i} = N\, ,\ee
such that the matrix $X$ has $N_{i}$ eigenvalues equal to $w_{i}$. In
particular, the generating function
\be\label{Rdef} R(z) = \Tr\frac{1}{z-X} =\sum_{k\geq
0}\frac{u_{k}}{z^{k+1}}\ee
is given by
\be\label{Rclopen} R(z) = \sum_{i}\frac{N_{i}}{z-w_{i}}\,\cdotp\ee

In the closed string description, we can use only the gauge invariant
operators $u_{k}$, not the matrix $X$. The equations of motion
\eqref{openclem} are then written as
\be\label{closedclem} \Tr \bigl(X^{n+1}W'(X)\bigr) = 0 = \sum_{k\geq
0}g_{k}u_{n+k+1}\, ,\quad n\geq -1\, .\ee
In terms of $R(z)$, this is equivalent to the existence of a
degree $d-1$ polynomial $\Delta$ such that
\be\label{clano} W'(z) R(z) = \Delta(z)\, .\ee
The vanishing of the terms proportional to negative powers of $z$ in
the large $z$ expansion of the left hand side of \eqref{clano} is
indeed equivalent to the equations \eqref{closedclem}. The most
general solution to \eqref{closedclem} or \eqref{clano} is given by
\be\label{Rclclosed} R(z) =\frac{\Delta(z)}{W'(z)}=
\sum_{i=1}^{d}\frac{c_{i}}{z-w_{i}}\,\cdotp\ee
The constants $c_{i}$ can be arbitrary complex numbers, with the only 
constraint
\be\label{cisum}\sum_{i=1}^{d}c_{i}= N\ee
that follows from the definition of $R(z)$. 

To make contact with the open string formula \eqref{Rclopen}, we have
to prove that the $c_{i}$ must be positive integers. This is obvious
in the open string framework since $c_{i}=N_{i}$ is then identified
with the number of eigenvalues of the matrix $X$ that are equal to
$w_{i}$. The question is: how can we understand this quantization
condition in a formulation where only the gauge invariant operators
$u_{k}$ are available?

The fundamental idea is to implement the constraints \eqref{defPp}
\cite{ferchiral}. We are going to show the simple\smallskip\\
\textbf{Theorem}. \emph{The equations \eqref{closedclem} are
consistent with the constraints \eqref{defPp} if and only if the
constants $c_{i}$s in \eqref{Rclclosed} are positive integers. In
particular, the integrals $\frac{1}{2i\pi}\oint\! R\,\d z$ over any
closed contours are integers.}\smallskip\\
This is a toy version of the chiral ring consistency theorem that we
shall prove later. Very concretely, it means that a set of variables
$u_{k}$ given by the formulas
\be\label{uktoy} u_{k} = \sum_{i=1}^{d}c_{i}w_{i}^{k}\ee
can satisfy the constraints \eqref{defPp} if and only if the $c_{i}$s
are positive integers. To prove this simple algebraic result, we use
the following trick. We introduce the function $F(z)$ defined by the
conditions
\be\label{Fdef}\frac{F'(z)}{F(z)} = R(z)\, ,\quad F(z)
\underset{z\rightarrow\infty}{\sim} z^{N}\, .\ee
In terms of the matrix $X$, one would simply have $F(z)=\det (z-X)$,
but we do not want to use the matrix $X$ here but only deal with the
gauge invariant variables $u_{k}$. The function $F$ is expressed in
terms of these variables by integrating \eqref{Rdef},
\be\label{Fexp} F(z) = z^{N}\exp\Bigl(-\sum_{k\geq
1}\frac{u_{k}}{kz^{k}}\Bigr)\, .\ee
The crucial algebraic property is that the relations \eqref{defPp} are
\emph{equivalent} to the fact that $F(z)$ is a polynomial. A very
effective way to compute the polynomials $P_{p}$ is actually to write
that the terms with a negative power of $z$ in the large $z$ expansion
of the right-hand side of \eqref{Fexp} must vanish. If $F$ is a
polynomial, then of course it is a single-valued function of $z$, and
thus
\be\label{qccl} \frac{1}{2i\pi}\oint\! R\,\d z = 
\frac{1}{2i\pi}\oint\d\ln F\in\mathbb Z\, .\ee
In particular, the $c_{i}$s are integers. They are positive because
$F$ does not have poles. Conversely, if the $c_{i}$ are positive
integers, then we can introduce the matrix $X$ defined to have $c_{i}$
eigenvalues equal to $w_{i}$ for all $i$. The relations \eqref{defPp}
are then automatically satisfied.

\subsection{The model}
\label{modelSec}

Our aim in the present paper is to generalize the above analysis to
the full non-perturbative quantum theory, by analysing the consistency
between the quantum versions of \eqref{closedclem} and \eqref{defPp}
to prove that the periods $\frac{1}{2i\pi}\oint\! R\,\d z$ must always
be quantized. These quantization conditions are highly non-trivial
constraints, known to be equivalent to a specific form of the
Dijkgraaf-Vafa glueball superpotential, including the
Veneziano-Yankielowicz coupling-independent part, and to contain the
crucial information on the non-perturbative dynamics of the theory in
the matrix model formalism \cite{CSW2,ferproof,ferchiral}.

We shall focus on the $\uN$ theory with one adjoint chiral superfield
$X$ and $\Nf$ flavours of fundamentals $(\tilde Q^{a},Q_{b})$. We
always assume that the theory is asymptotically free or conformal in
the UV,
\be\label{Nfconst}\Nf\leq 2N\, .\ee
When $\Nf<2N$ the instanton factor is given by
\be\label{inst1} q = \La^{2N-\Nf}\ee
in terms of the dynamically generated complex scale $\La$. When
$\Nf=2N$ we have
\be\label{inst2} q = e^{-8\pi^{2}/g^{2}+ i \vartheta}\ee
in terms of the Yang-Mills coupling constant $g$ and $\vartheta$
angle. The tree-level superpotential has the form
\be\label{treegen} \wt = \Tr W(X) +
\sum_{1\leq a,b\leq\Nf}{}^{T}\tilde Q^{a}m_{a}^{\
b}(X)Q_{b}\, .\ee
The derivative of $W(z)$ is as in \eqref{wtdef}, and $m_{a}^{\ b}(z)$
is a $\Nf\times\Nf$ matrix-valued polynomial,
\be\label{mabkdef} m_{a}^{\ b}(z) = \sum_{k= 0}^{\delta} m_{a,\, k}^{\
b}\, z^{k}\, ,\ee
with
\be\label{detm}\det m(z) = U(z) = U_{0}\prod_{Q=1}^{\Nf\delta}(z-b_{Q})\,
.\ee
It is useful to introduce the symmetric polynomials
\be\label{sigmadef} \sigma_{\alpha} = \sum_{Q_{1}<\cdots<Q_{\alpha}}
b_{Q_{1}}\cdots b_{Q_{\alpha}}\, ,\quad 1\leq \alpha\leq\Nf\delta\,
.\ee
We shall consider the case where $m_{a}^{\ b}(z)$ is a linear function
of $z$,
\be\label{deltaone}\delta=1\, ,\ee
in the following.\footnote{This is not strictly necessary as long as
the constraint $\Nf\delta\leq 2N$ is satisfied.} 

The classical theory has a large number of vacua obtained by
extremizing the superpotential \eqref{treegen}. The most general
solution $|N_{i};\nu_{Q}\ranglecl$ is labeled by the numbers of
eigenvalues of the matrix $X$, $N_{i}\geq 0$ and $\nu_{Q}=0$ or $1$,
that are equal to $w_{i}$ and $b_{Q}$ respectively \cite{CSW2}. The
constraint
\be\label{trivialconst}\sum_{i=1}^{d}N_{i} + \sum_{Q=1}^{\Nf}\nu_{Q} =
N\ee
must be satisfied. The gauge group $\uN$ is broken down to
$\text{U}(N_{1})\times\cdots\times\text{U}(N_{d})$ in a vacuum
$|N_{i};\nu_{Q}\ranglecl$. We shall call the number of non-zero
integers $N_{i}$ the \emph{rank} $r$ of the vacuum.

In addition to \eqref{ukdef}, we have other basic gauge invariant
operators in the theory that are constructed by using the vector
chiral superfield $W^{\alpha}$,
\be\label{opedef} u_{k}^{\alpha} = \frac{1}{4\pi}\Tr W^{\alpha}X^{k}\,
,\quad v_{k} = -\frac{1}{16\pi^{2}}\Tr W^{\alpha}W_{\alpha}X^{k}\,
,\quad w_{a,\, k}^{\ b} = {}^{T}\tilde Q^{b}X^{k}Q_{a}\, .\ee
The associated generating functions are defined by
\be\label{genfgen} \mathcal W^{\alpha}(z) = \sum_{k\geq
0}\frac{u_{k}^{\alpha}}{z^{k+1}}\,\cvp\quad S(z) = \sum_{k\geq
0}\frac{v_{k}}{z^{k+1}}\,\cvp\quad G_{a}^{\ b}(z) = \sum_{k\geq
0}\frac{w_{a,\, k}^{\ b}}{z^{k+1}}\,\cdotp\ee

The relations that replace \eqref{closedclem}, or equivalently
\eqref{clano}, in the full quantum theory are given by the following
generalized Konishi anomaly equations \cite{seifla}
\begin{gather}\label{a1} NW'(z)R(z) + N{m'}_{a}^{\ b}(z)G_{b}^{\ a}(z) -
2 S(z)R(z) - 2 \mathcal W^{\alpha}(z)\mathcal W_{\alpha}(z) =
\Delta_{R}(z) \\
\label{a2} NW'(z)\mathcal W^{\alpha}(z) - 2 S(z)\mathcal W^{\alpha}(z)
= \Delta^{\alpha}(z)\\\label{a3}
NW'(z) S(z) - S(z)^{2} = \Delta_{S} (z) \\\label{a4}
NG_{a}^{\ c}(z) m_{c}^{\ b}(z) - S(z)\delta_{a}^{b} =
\Delta_{a}^{b}(z)\\\label{a5}
Nm_{a}^{\ c}(z) G_{c}^{\ b}(z) - S(z)\delta_{a}^{b} =
\tilde\Delta_{a}^{b}(z)\, .
\end{gather}
The functions $\Delta_{R}$, $\Delta^{\alpha}$, $\Delta_{S}$,
$\Delta_{a}^{b}$ and $\tilde\Delta_{a}^{b}$ must be polynomials. By
expanding \eqref{a1}--\eqref{a5} at large $z$, and writing that the
terms proportional to negative powers of $z$ must vanish, we obtain an
infinite set of constraints on the gauge invariant operators, valid
for any integer $n\geq -1$,
\begin{gather}\label{a1b}N\sum_{k\geq 0}\bigl(g_{k}u_{n+k+1} +
(k+1)m_{a,\, k+1}^{\ b}w_{b,\, n+k+1}^{\ a}\bigr) -
2\sum_{k_{1}+k_{2}=n}\bigl(u_{k_{1}}v_{k_{2}} +
u_{k_{1}}^{\alpha}{u_{k_{2}}}_{\alpha}\bigr) = 0\\ \label{a2b}
N\sum_{k\geq 0} g_{k}u_{n+k+1}^{\alpha} - 2\sum_{k_{1}+k_{2}=n}
v_{k_{1}}u_{k_{2}}^{\alpha} = 0\\ \label{a3b} N\sum_{k\geq 0}
g_{k}v_{n+k+1} - \sum_{k_{1}+k_{2}=n} v_{k_{1}}v_{k_{2}} = 0\\
\label{a4b} N\sum_{k\geq 0} w_{a,\, k+n+1}^{\ c}m_{c,\, k}^{\ b} -
v_{n+1}\delta_{a}^{b} = 0\\ \label{a5b} N\sum_{k\geq 0}m_{a,\, k}^{\
c}w_{c,\, k+n+1}^{\ b} - v_{n+1}\delta_{a}^{b} = 0\, .
\end{gather}
\begin{figure}
\centerline{\epsfig{file=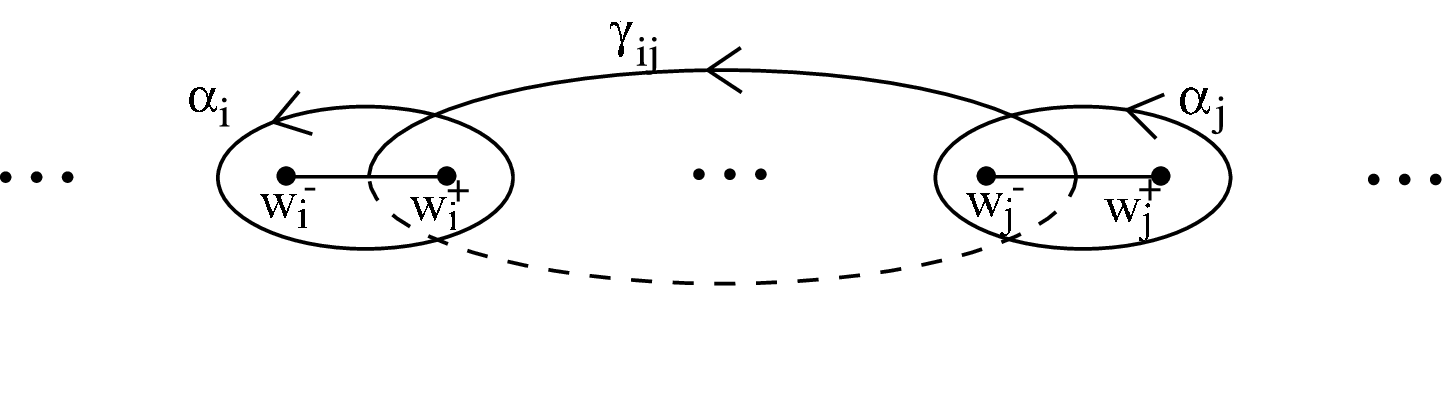,width=14cm}}
\caption{The hyperelliptic Riemann surface $\mathcal C$,
with the contours $\alpha_{i}$ and $\gamma_{ij}$ used in
the main text.
\label{fig1}}
\end{figure}

Equations \eqref{a1}--\eqref{a5} show that the generating functions
are meromorphic functions on a hyperelliptic Riemann surface of the
form
\be\label{Cdef}\mathcal C_{r}:\ y_{r}^{2} =
\prod_{i=1}^{r}(z-w_{i}^{-})(z-w_{i}^{+})\, .\ee
The integer $r$, called the rank of the solution, must satisfy
\be\label{rcond} r\leq d\ee
and we have
\be\label{factcond} W'(z)^{2} - 4\Delta_{d-1}(z) =
\phi_{d-r}(z)^{2}y_{r}^{2}\ee
for some polynomials $\Delta_{d-1}=\Delta_{S}/N^{2}$ and $\phi_{d-r}$
of degrees $d-1$ and $d-r$ respectively. The curve \eqref{Cdef} with
some closed contours is depicted in Figure 1. It corresponds to the
geometry of the closed string background. We shall use extensively in
the following the most general solution to \eqref{a1}--\eqref{a5} of
rank $r$ for the expectation value $\langle R(z)\rangle_{r}$. It has
the form \cite{CSW2}
\be\label{gensolR} \bigl\langle R(z)\bigr\rangle_{r} =
\frac{C_{r-1}}{y_{r}} + \frac{1}{2}\frac{U'}{U} -
\frac{1}{2y_{r}}\sum_{Q}\frac{(1-2\nu_{Q})y_{r}(z=b_{Q})}{z-b_{Q}}\,\cdotp
\ee
The polynomial $C_{r-1}=\frac{1}{2} (2N-\Nf)z^{r-1}+\cdots$ is of
degree $r-1$ and is a priori unknown except for its term of highest
degree that is fixed by the large $z$ asymptotics of $R(z)$. In the
classical limit, the solutions \eqref{gensolR} correspond to the rank
$r$ classical vacua $|N_{i};\nu_{Q}\ranglecl$ described previously.
Quantum mechanically, the anomaly equations \eqref{gensolR} leave
$2r-1$ arbitrary parameters, which are the coefficients of $C_{r-1}$
and of $\Delta_{d-1}$ that are not fixed by the factorization
condition \eqref{factcond}. These unknown parameters are the quantum
analogues of the coefficients $c_{i}$ in \eqref{Rclclosed}, and our
main goal is to show that they are fixed by a quantum version of the
simple consistency proof explained in \ref{classexSec}.

\subsection{Chiral ring relations and anomaly equations}
\label{NPASec}

The model \eqref{treegen} has a useful $\text{SU}(\Nf)_{\text
L}\times\text{SU}(\Nf)_{\text R}\times \u_{\text A}\times\u_{\text
B}\times\u_{\text R}$ global symmetry. The charges of the various
parameters and operators of the theory are given in the following
table
\be\label{asign}
\hskip 1cm\begin{matrix}
& u_{k} & u_{k}^{\alpha} & v_{k} & w_{a,\, k}^{\ b} & g_{k} & 
m_{a,\, k}^{\ b} &  \sigma_{k} & U_{0} & q \\
{\rm U}(1)_{\rm A} & k & k & k & k-1 & -k-1 & -k+1 & k & 0 & 2N-\Nf \\
{\rm U}(1)_{\rm B} & 0 & 0 & 0 & 2 & 0 & -2 & 0 & -2\Nf & 2\Nf \\
{\rm U}(1)_{\rm R} & 0 & 1 & 2 & 2 & 2 & 0 & 0 & 0 & 
0\\
\text{SU}(\Nf)_{\text L}& \mathbf 1& \mathbf 1& \mathbf 1&
\boldsymbol{N_{\text f}}&\mathbf 1 &\boldsymbol{N_{\text f}} &\mathbf
1 &\mathbf 1 &\mathbf 1 \\
\text{SU}(\Nf)_{\text R}& \mathbf 1& \mathbf 1& \mathbf 1&
\overline{\boldsymbol{N_{\text f}}}&\mathbf 1
&\overline{\boldsymbol{N_{\text f}}} &\mathbf 1 &\mathbf 1
&\hphantom{,\,}\mathbf 1 \, . \end{matrix}\ee
It is useful for our purposes to consider the subring $\mathscr A_{0}$
of the chiral ring of the theory that is invariant under
$\text{SU}(\Nf)_{\text L}\times\text{SU}(\Nf)_{\text R}\times\u_{\text
B}\times\u_{\text R}$. This subring is generated by the operators
$u_{k}$ and the parameters\footnote{It is convenient to include the
parameters, which can always be promoted to background chiral
superfields, in the chiral ring.} $\sigma_{k}$ and
\be\label{defqspe}\q = U_{0}q\, .\ee
It is a simple polynomial ring given by
\be\label{A0}\mathscr A_{0} = \mathbb
C[\q,\sigma_{1},\ldots,\sigma_{\Nf}, u_{1},\ldots,u_{N}]\, .\ee
As stressed in the Section 2 of \cite{ferchiral}, a polynomial ring
has no deformation, and thus \eqref{A0}, which is trivially valid at
the classical level due to the relations \eqref{defPp}, is also valid
in the full quantum theory. The meaning of this statement is simply
that any operator in $\mathscr A_{0}$ can be expressed as a finite sum
of finite products of $U_{0}q$, $\sigma_{k}$ for $1\leq k\leq\Nf$ and
$u_{k}$ for $1\leq k\leq N$, a rather trivial result. It is sometimes
claimed in the literature that the ring $\mathscr A_{0}$ is deformed
because the relations \eqref{defPp} can get quantum corrections. This
is not correct. In the full quantum theory, we can \emph{define} what
is meant by $u_{k}\in\mathscr A_{0}$ for $k>N$ by a relation of the
form
\be\label{defPquantum} u_{N+p} = \mathscr
P_{p}(u_{1},\ldots,u_{N},\sigma_{1},\ldots,\sigma_{\Nf},\q)\, ,\quad
p\geq 1\, . \ee
The $\mathscr P_{p}$ are chosen to be consistent with the symmetries
\eqref{asign} and the classical limit \eqref{defPp}, but can be
completely arbitrary otherwise. It can be convenient to work with a
particular definition \eqref{defPquantum}, and we shall see shortly
that there is indeed a canonical choice, but this remains a choice and
has no physical content \cite{ferchiral}. Let us note that a parallel
discussion applies to the variables $u_{k}^{\alpha}$, $v_{k}$ and
$w_{a,\, k}^{\ b}$, which are independent only for $0\leq k\leq N-1$.

The equations \eqref{a1}--\eqref{a5} were derived in perturbation
theory in \cite{Ano,seifla}. At the non-perturbative level, these
equations \emph{do} get quantum corrections. However, these quantum
corrections have a very special form. The general theorem is as
follows:\smallskip\\
\textbf{Non-perturbative anomaly theorem} \cite{mic3}:
\emph{The non-perturbative corrections to the generalized anomaly
equations are such that they can be absorbed in a non-perturbative
redefinition of the variables that enter the equations.}\smallskip\\
This means that there exists a canonical choice for the definitions of
the variables $u_{k}$ for $k>N$ as in \eqref{defPquantum}, and other
similar canonical definitions of the variables $u_{k-1}^{\alpha}$,
$v_{k-1}$ and $w_{a,\, k-1}^{\ b}$ for $k>N$, that make all the
non-perturbative corrections \emph{implicit}. The theorem has been
proven recently in the case of the theory with no flavour \cite{mic3}.
The arguments used in \cite{mic3} can in principle be generalized
straightforwardly, and we shall take the result for granted in the
theory with flavours as well.

\subsection{The chiral ring consistency theorem}
\label{crctSec}

We can now state the quantum version of the classical problem solved
in Section \ref{classexSec}. On the one hand, in the closed string
description, the theory is described by the equations
\eqref{a1}--\eqref{a5} or equivalently by \eqref{a1b}--\eqref{a5b}. On
the other hand, we know that the existence of the open string
formulation implies that relations of the form \eqref{defPquantum}
must exist. These relations imply that there are only a finite number
of independent variables. The anomaly equations
\eqref{a1b}--\eqref{a5b} thus yield an infinite set of constraints on
a finite set of independent variables. Generically, such an
overconstrained system of equations is inconsistent. The main result
of the present work is to prove the\smallskip\\
\textbf{Chiral ring consistency theorem}: \emph{The system of
equations \eqref{a1b}--\eqref{a5b} is consistent with the existence of
relations of the form \eqref{defPquantum} if and only if the periods
of the gauge theory resolvent $\frac{1}{2i\pi}\oint\! R\,\d z$ are
integers. The relations \eqref{defPquantum} (and all the other
relations amongst chiral operators) are then fixed in a unique
way.}\smallskip\\
This theorem was conjectured in \cite{ferchiral}. As discussed in
\ref{modelSec}, the equations \eqref{a1b}--\eqref{a5b} imply that $R$
is a meromorphic function on a hyperelliptic curve of the form
\eqref{Cdef}. The theorem then states that the algebraic consistency
of the chiral ring implies
\begin{align}\label{QC1}&\frac{1}{2i\pi}\oint_{\alpha_{i}}\! \langle
R\rangle_{r}\,\d z\in\mathbb Z\\
\label{QC2}&\frac{1}{2i\pi}\oint_{\gamma_{ij}}\! \langle
R\rangle_{r}\,\d z\in\mathbb Z\, ,\end{align}
where the contours $\alpha_{i}$ and $\gamma_{ij}$ are defined in
Figure 1. Actually, the $\alpha_{i}$-periods are automatically
positive, as we shall see. Several comments on this result are in
order.

First, the equations \eqref{QC1} and \eqref{QC2} yield $2r-2$
non-trivial constraints on \eqref{gensolR} (the sum of the equations
\eqref{QC1} is trivial because the asymptotic condition
\be\label{Rasy} \bigl\langle
R(z)\bigr\rangle\underset{z\rightarrow\infty}{\sim}\frac{N}{z}\ee
is automatically satisfied by \eqref{gensolR}). Thus the solution is
uniquely fixed, up to a single unknown that can be identified with the
quantum deformation parameter. We shall explain in \ref{rankoneSec}
how to relate precisely this parameter to the instanton factor $q$.

The quantization conditions \eqref{QC1} are the quantum versions of
the classical result \eqref{qccl}. Note that it is not correct to
claim that \eqref{QC1} is obvious because the period integral yields
the number of eigenvalues of the matrix $X$ in the cut
$[w_{i}^{-},w_{i}^{+}]$. This interpretation is completely erroneous
is the context of the finite $N$ gauge theory \cite{ferchiral}.
Actually, most of the possible definitions \eqref{defPquantum} would
violate \eqref{QC1}. The correct interpretation of equation
\eqref{QC1} is that it yields non-trivial constraints on the canonical
definitions of the variables for which the anomaly equations have the
simple form \eqref{a1b}--\eqref{a5b}, in line with the
non-perturbative anomaly theorem in Section \ref{NPASec}.

The quantization conditions \eqref{QC2} have no classical counterpart.
In the closed string formulation of the theory, three-form fluxes are
turned on. The associated flux superpotential coincides with the
Dijkgraaf-Vafa glueball superpotential $W_{\text{DV}}(S_{i})$, and the
equations \eqref{QC2} are equivalent to the extremization of
$W_{\text{DV}}$ \cite{CSW2}. The chiral ring consistency theorem
thus answers, for the chiral sector of the theory, the questions asked
in Section 1: the existence of the relations \eqref{defPquantum},
which are trivial off-shell identities in the open string description,
are seen in the closed string formulation only after implementing the
closed string dynamical equations of motion. The exchange of off-shell
identities and on-shell dynamical equations in the open/closed string
duality was emphasized in \cite{mic3}.

Another important consequence of the theorem is to lift the mystery of
the Veneziano-Yankielowicz term $f(S_{i})$ in $W_{\text{DV}}$. It was
shown in \cite{Ano} that an arbitrary function $f(S_{i})$, depending
on the glueball superfields $S_{i}$ but independent of the couplings
in the tree-level superpotential \eqref{treegen}, could be added to
$W_{\text{DV}}$ without spoiling the correspondence with the matrix
model. The term $f(S_{i})$ plays of course a crucial r\^ole in fixing
the on-shell values of the glueballs, and is at the heart of the
non-perturbative gauge dynamics. However, it is left unconstrained by
the anomaly equations \eqref{a1}--\eqref{a5}, whose most general
solutions are simply parametrized by the $S_{i}$. From the point of
view of the matrix model, the glueballs $S_{i}$ are identified with
the filling fractions which are completely arbitrary parameters. For
these reasons, and as discussed at length in \cite{Ano} for example,
the determination from first principles of the function $f(S_{i})$
seemed to be out of reach. We now see that the situation is
conceptually must simpler that what might have been expected
\cite{ferchiral}: the filling fractions, and thus the
Veneziano-Yankielowicz term $f(S_{i})$, are fixed entirely by imposing
the consistency between \eqref{a1}--\eqref{a5} and
\eqref{defPquantum}. The fact that this term ought to be fixed by
general consistency conditions was first emphasized in
\cite{ferproof}.

To prove our main theorem, we are going to show that the function $F$
defined by \eqref{Fdef} must satisfy the fundamental equation
\be\label{Feq} F(z) + \frac{qU(z)}{F(z)} = H(z)\ee
for a polynomial $H = (1+\q\delta_{N_{\text f},2N})z^{N}+\cdots$ of
degree $N$. In the classical theory $q=0$, \eqref{Feq} simply says
that $F$ must be a polynomial, and we have explained after
\eqref{Fexp} that this condition is equivalent to the relations
\eqref{defPp}. Similarly, in the quantum theory, \eqref{Feq} is
\emph{equivalent} to a particular quantum corrected form
\be\label{rel}u_{N+p} = \mathscr
P_{p}^{(0)}(u_{1},\ldots,u_{N},\sigma_{1},\ldots,\sigma_{\Nf},\q)\,
,\quad p\geq 1\, ,\ee
for the relations \eqref{defPquantum} \cite{ferchiral}. This result is
obtained straightforwardly by expanding the left hand side of
\eqref{Feq} at large $z$.

Equation \eqref{Feq} implies that
\be\label{Fform} F = \frac{1}{2}\Bigl( H + \sqrt{H^{2} - 4 q 
U}\Bigr)\ee
is a meromorphic function on the hyperelliptic surface 
\be\label{Ydef}\tilde{\mathcal C}:\ Y^{2} = H(z)^{2} - 4qU(z)\, . \ee
The generating function
\be\label{Rform} R=\frac{F'}{F} = \frac{1}{2}\frac{U'}{U} + \Bigl(H' -
\frac{U'H}{2U}\Bigr)\frac{1}{\sqrt{H^{2} - 4qU}} \ee
is then also automatically a meromorphic function on the same curve
$\tilde{\mathcal C}$. From the single-valuedness of $F$ on
$\tilde{\mathcal C}$, we deduce that
\be\label{qcquantum}\frac{1}{2i\pi}\oint_{c} R\,\d z =
\frac{1}{2i\pi}\oint_{c}\d\ln F\in\mathbb Z\ee
for any closed contour $c$. Note that the consistency of \eqref{Rform}
with the fact that $\langle R\rangle_{r}$ must be well-defined on
the curve \eqref{Cdef} implies that the following factorization
condition must hold in the rank $r$ vacua
\be\label{fact2} \langle H(z)\rangle_{r}^{2} - 4 q U(z) =
\psi_{N-r}(z)^{2}y_{r}^{2}\, ,\ee
for some degree $N-r$ polynomial $\psi_{N-r}$. The equations
\eqref{qcquantum} thus automatically imply \eqref{QC1} and
\eqref{QC2}. The positivity of the $\alpha_{i}$-periods is a direct
consequence of the classical limit. We shall focus on proving
\eqref{Feq} in the following.

\section{The proof of the main theorem}
\setcounter{equation}{0}
\subsection{Generalities}

We suppose from now on that $\Nf=2N$. The other cases with $\Nf<2N$
can be obtained by integrating out some flavours, sending their masses
to infinity. If not explicitly stated otherwise, we shall always
assume that the degree of $W'$ in \eqref{wtdef} is
\be\label{dN} d = N\, .\ee
This is not a restriction, because the $\u_{\text R}$ symmetry implies
that the relations \eqref{defPquantum} we want to study cannot depend
on the couplings $g_{k}$ in $W$.

It is convenient to define new variables $x_{1},\ldots,x_{N}$ by the
relations
\be\label{xidef} u_{k} = \sum_{i=1}^{N}x_{i}^{k}\quad\text{for}\ 1\leq
k\leq N\, .\ee
Strictly speaking, the $x_{i}$s are not in the chiral subring
$\mathscr A_{0}$, but they can always be introduced by using the
following algebraic trick. We consider the polynomial part $F_{0}$ of
the function $F$ at large $z$,
\be\label{F0def} F(z) = F_{0}(z) + \mathcal O(1/z)\, .\ee
Equation \eqref{Fexp} shows that the coefficients of $F_{0}$ are
themselves polynomials in the $u_{k}$, and thus $F_{0}\in\mathscr
A_{0}[T]$ where $T$ is an undeterminate. It is then trivial to check
that the $x_{i}$ satisfying \eqref{xidef} are the roots of the
polynomial $F_{0}$ in its splitting field.\footnote{The existence of
the splitting field for any polynomial and thus of the variables
$x_{i}$ is ensured by standard theorems in elementary algebra, see for
example \cite{Lang}.} 

We can use the new variables to rewrite the relations
\eqref{defPquantum} in the form
\be\label{relQ} u_{N+p} = \mathscr
P_{p}(\x,\b,\q)\, .\ee
We use boldface letters to represent collectively a set of variables,
for example $\x$ represents all the $x_{i}$, $1\leq i\leq N$. In
\eqref{relQ}, $\mathscr P_{p}$ must be invariant under the action of
the permutation group $S_{N}\times S_{2N}$ that act on the $x_{i}$s
and the $b_{Q}$s independently.\footnote{The $\mathscr P_{p}$
appearing in \eqref{defPquantum} and \eqref{relQ} are of course not
the same. We use the same notation because they coincide when the
relations \eqref{xidef} and \eqref{sigmadef} are taken into account.}

Let us introduce the vector spaces $\mathcal V_{k}$ of arbitrary power
series in $\x$, $\b$ and $\q$ that are invariant under the action of
$S_{N}\times S_{2N}$ and that are homogeneous of degree $k$, the
degree being identified with the A-charge defined by \eqref{asign}.
Clearly, $\mathscr P_{p}\in\mathcal V_{N+p}$. Since $\x$ and $\b$ are
of degree one, elements of $\mathcal V_{k}$ must be polynomials in the
$x_{i}$s and $b_{Q}$s. On the other hand, $\q$ is of degree $2N-\Nf$.
In the general case $\Nf=2N$ we are considering, arbitrary powers of
$\q$ can thus in principle appear.

The equations \eqref{relQ} are operator equations. Taking the
expectation value, we get
\be\label{relvev}\langle u_{N+p}\rangle = \mathscr
P_{p}\bigl(\langle\x\rangle,\b,\q\bigr)\, ,\ee
in \emph{all} the vacua of the theory. Of course, $\langle
u_{N+p}\rangle$ and $\langle\x\rangle$ depend on the particular vacua
under consideration, but the polynomials $\mathscr P_{p}$ do
\emph{not}. With this constraint, it is easy to realize that the
equations \eqref{relvev} cannot be consistent with the most general
solution \eqref{gensolR} to the anomaly equations. We are going to
prove that consistency is achieved only when \eqref{Feq} is satisfied,
which corresponds to the quantization conditions \eqref{QC1} and
\eqref{QC2} and to the particular form \eqref{rel}
\be\label{PP0id}\mathscr P_{p}=\mathscr P_{p}^{(0)}\ee
of the relations \eqref{relvev}.

We shall use the following strategy. The solutions \eqref{gensolR} are
uniquely fixed for the rank zero vacua. We are going to show that this
implies that the polynomial $\mathscr P_{1}$ must be equal to
$\mathscr P_{1}^{(0)}$. But this provides a non-trivial operator
constraint, that fixes uniquely the solutions \eqref{gensolR} at rank
one. Analysing the form of these solutions, we can then show that
\eqref{PP0id} or \eqref{rel} must be valid at least for $1\leq p\leq
N+5$. This yields $N+5$ operator constraints, that can be used to fix
the solutions \eqref{gensolR} for $2r-1\leq N+5$. In particular, we
know the rank two solutions for any $N$. This turns out to imply that
\eqref{rel} must be true at least for $p\leq 2N+7$. This yields a
total of $2N+7$ constraints, a number greater than the maximum number
of unknown parameters $2N-1$ that can appear in \eqref{gensolR}. We
can then check that all the resulting solutions do satisfy \eqref{rel}
for all $p$.

\subsection{Using the rank zero vacua}
\label{rank0Sec}

Let us start by looking at the vacua of rank zero, that correspond to
a completely broken gauge group. There is no free parameter in this
case, and thus the solution must be completely fixed. This is not
difficult to check. The factorization condition \eqref{factcond}
yields when $r=0$
\be\label{fact0} 4\Delta_{N-1} = W'^{2} - \phi_{N}^{2} =
(W'-\phi_{N})(W'+\phi_{N})\, .\ee
Since $W'$ and $\phi_{N}$ are of degree $N$, whereas $\Delta_{N-1}$ is
of degree $N-1$, \eqref{fact0} implies that $\phi_{N}=\pm W'$ and
$\Delta_{N-1}=0$. Equation \eqref{a3} then shows that $\langle
S(z)\rangle_{r=0}= 0$. From the other anomaly equations, we
immediately derive that there can be no quantum correction at all. In
particular,
\be\label{R0}\bigl\langle R(z)\bigr\rangle_{r=0} =
\sum_{Q=1}^{2N}\frac{\nu_{Q}}{z-b_{Q}}\,\cvp\quad \bigl\langle
F(z)\bigr\rangle_{r=0} = \prod_{Q=1}^{2N}(z-b_{Q})^{\nu_{Q}}\, .\ee
We can thus compute
\be\label{Feq0} \langle F(z)\rangle_{r=0} + \frac{qU(z)}{\langle
F(z)\rangle_{r=0}} = \prod_{Q=1}^{2N}(z-b_{Q})^{\nu_{Q}} +
\q\prod_{Q=1}^{2N}(z-b_{Q})^{1-\nu_{Q}}\, ,\ee
which is indeed a polynomial, consistently with \eqref{Feq}. This shows
that
\be\label{relvev0}\langle u_{N+p}\rangle_{r=0} = \mathscr
P_{p}^{(0)}\bigl(\langle\x\rangle_{r=0},\b,\q\bigr)\, ,\quad p\geq 1\,
,\ee
in the vacua of rank zero.

What can we learn from \eqref{relvev0} on the possible forms of the
operator relations \eqref{relQ}? Let us decompose $\mathscr P_{p}$ as
the sum of two terms,
\be\label{reldec}u_{N+p} = \mathscr P_{p}^{(0)}(\x,\b,\q) + \mathscr
P_{p}^{(1)}(\x,\b,\q)\, .\ee
Equation \eqref{relvev0} is equivalent to the constraints
\be\label{P1vanish}\mathscr
P_{p}^{(1)}\bigl(\langle\x\rangle_{r=0},\b,\q\bigr) = 0\, ,\quad p\geq
1\, ,\ee
where the expectation values are taken in the rank zero vacua only. It
is extremely important to understand that this constraint does
\emph{not} imply that $\mathscr P_{p}^{(1)}=0$, because the variables
$x_{i}$ and $b_{j}$ are not algebraically independent in the rank zero
vacua. This subtlety is completely general. When one focuses on a
special set of vacua, operator relations can only be determined modulo
the ideal generated by the exceptional chiral ring relations that are
valid only in the particular vacua under consideration. It is clear
that there are many such relations in the rank zero vacua. Actually,
the fact that there are no quantum corrections in \eqref{R0} ensures
that the classical relations \eqref{defPp} must be valid,
\be\label{Ppclzero} \langle u_{N+p}\rangle_{r=0} = P_{p}\bigl(\langle
x\rangle_{r=0}\bigr)\,\quad p\geq 1\, .\ee
Consistency with \eqref{relvev0} implies that
\be\label{zerorel}\mathscr
P_{p}^{(0)}\bigl(\langle\x\rangle_{r=0},\b,\q\bigr) -
P_{p}\bigl(\langle\x\rangle_{r=0}\bigr) = 0\, .\ee
The relations \eqref{zerorel} prevent an analysis based only on the
rank zero vacua to fix unambiguously the operator relations
\eqref{reldec}.

Even though \eqref{P1vanish} does not show that $\mathscr P_{p}^{(1)}$
must vanish for all $p$, it does put some non-trivial constraints. In
the rank zero vacua, \eqref{R0} shows that the set
$\{x_{1},\ldots,x_{N}\}$ is identified with a subset of
$\{b_{1},\ldots,b_{2N}\}$. Taking into account the fact that $\mathscr
P^{(1)}$ is symmetric under permutation of the $b_{i}$s, we deduce
that \eqref{P1vanish} is equivalent to
\be\label{P1v02} \mathscr
P^{(1)}_{p}(x_{1}=b_{1},x_{2}=b_{2},\ldots,x_{N}=b_{N},\b,\q) = 0\,
.\ee
Let us now show the following\smallskip\\
\textbf{Proposition}: \emph{Let $A\in\mathcal V_{n}$, $A\not = 0$,
such that}
\be\label{constA} A(x_{1}=b_{1},x_{2}=b_{2},\ldots,x_{N}=b_{N},\b,\q)
= 0\, .\ee
\emph{Then $\deg A = n\geq N+1$. Moreover, if $\deg A = N+1$, then $A$
must be of the form}
\be\label{AN1} A(\x,\b,\q) = \mathsf a
(\q)\sum_{i=1}^{N}\frac{\prod_{Q=1}^{2N}(x_{i}-b_{Q})}{\prod_{j\not =
i}(x_{i}-x_{j})}\,\cvp\ee
\emph{for some $\x$- and $\b$-independent power series $\mathsf a$ in
$\q$.}\smallskip\\
To prove the proposition, we note that \eqref{constA} ensures that
there exists $k_{0}\leq N$, defined to be the smallest integer such
that $A(x_{1}=b_{1},\ldots,x_{k_{0}}=b_{k_{0}},x_{k_{0}+1},
\ldots,x_{N},\b,\q)$ identically vanishes (for all
$x_{k_{0}+1},\ldots,x_{N}$). By symmetry of the $b_{i}$s, we have
\be\label{Aded0}A(x_{1}=b_{1},\ldots,x_{k_{0}-1}=b_{k_{0}-1},
x_{k_{0}}=b_{k},x_{k_{0}+1}, \ldots,x_{N},\b,\q) = 0\ \text{for\ all}\
k\geq k_{0}\, . \ee
Seeing $A(x_{1}=b_{1},\ldots,x_{k_{0}-1}=b_{k_{0}-1},x_{k_{0}},
\ldots,x_{N},\b,\q)$ as a polynomial of a single variable $x_{k_{0}}$,
\eqref{Aded0} implies that
\be\label{Aded1}
A(x_{1}=b_{1},\ldots,x_{k_{0}-1}=b_{k_{0}-1},x_{k_{0}},
\ldots,x_{N},\b,\q) =\prod_{k=k_{0}}^{2N}(x_{k_{0}}-b_{k})B \ee
for some non-zero $B$ ($B=0$ would contradict the defining property of
$k_{0}$). In particular,
\be\label{degC1} \deg
A(x_{1}=b_{1},\ldots,x_{k_{0}-1}=b_{k_{0}-1},x_{k_{0}},
\ldots,x_{N},\b,\q)\geq 2N-k_{0} +1\, .\ee
Because $A(x_{1}=b_{1},\ldots,x_{k_{0}-1}=b_{k_{0}-1},x_{k_{0}},
\ldots,x_{N},\b,\q)$ does not vanish, and $A$ is homogeneous, we must
have $\deg A = \deg
A(x_{1}=b_{1},\ldots,x_{k_{0}-1}=b_{k_{0}-1},x_{k_{0}},
\ldots,x_{N},\b,\q)$ and thus $\deg A\geq 2N-k_{0}+1$. Since
$k_{0}\leq N$, we get $\deg A\geq N+1$, and the first part of the
proposition is proven.

It is not difficult to construct polynomials satisfying \eqref{constA}
for any degree $\geq N+1$ using the following trick. We consider the
rational function
\be\label{defRat} \rho(z) =
\frac{\prod_{Q=1}^{2N}(z-b_{Q})}{\prod_{i=1}^{N}(z-x_{i})}\,\cdotp\ee
At large $z$, we can expand
\be\label{rholargez} \rho(z) = z^{N} + \sum_{n\geq 1}
A_{n}(\x,\b)z^{N-n}\ee
in terms of $A_{n}\in\mathcal V_{n}$. Clearly, $\rho$ is a polynomial
if and only if the set $\{x_{1},\ldots,x_{N}\}$ is included in the set
$\{b_{1},\ldots,b_{2N}\}$. This is true if and only if all the terms
with a negative power of $z$ in the expansion \eqref{rholargez}
vanish, and thus the polynomials $A_{n}$ satisfy the constraint
\eqref{constA} for $n\geq N+1$. It is easy to check that the
polynomial $A_{N+1}$ is proportional to the right-hand side of
\eqref{AN1}. To show the uniqueness of the solution at degree $N+1$,
we note that \eqref{Aded1} implies that the coefficient of
$\sum_{i}x_{i}^{N+1}$ in a non-vanishing solution of degree $N+1$ must
be non-zero. If we have two non-zero solutions, we can always consider
a linear combination for which the terms in $\sum_{i}x_{i}^{N+1}$
cancel. The linear combination must then vanish, showing that the two
solutions we started with are proportional to each other. This ends
the proof of the proposition.

Using \eqref{P1v02}, we can apply the proposition to $A=\mathscr
P_{p}^{(1)}$. First, it shows that either $\mathscr P_{p}^{(1)}=0$ or
$\deg \mathscr P_{p}^{(1)}\geq N+1$. In the present case, this is
quite useless because $\deg\mathscr P_{p}^{(1)}=N+p\geq N+1$ is true
by construction. However, similar non-trivial inequalities will be of
great help in the next two subsections. Second, the proposition
implies that
\be\label{uN1rel}u_{N+1}= \mathscr P_{1}(\x,\b,\q) =
\sum_{i=1}^{N}x_{i}^{N+1} + \mathsf
a(\q)\sum_{i=1}^{N}\frac{\prod_{Q=1}^{2N}(x_{i}-b_{Q})}{\prod_{j\not =
i}(x_{i}-x_{j})}\,\cvp\ee
for some a priori unknown series $\mathsf a(\q) = a_{1}\q + \cdots$ in
$\q$. There are two possible attitudes with regard to the function
$\mathsf a(\q)$. A first possibility is to consider $\mathsf a$ to be
the quantum deformation parameter instead of $\q$. In particular,
expressing the results in terms of $\mathsf a$ instead of $\q$ is
irrelevant for the proof of the chiral ring consistency theorem and of
\eqref{QC1} and \eqref{QC2}. A second possibility is to insist on
using the instanton factor $\q$. It will be explained in subsection
\ref{rankoneSec} how to prove that
\be\label{aqrel} \mathsf a(\q) = (N+1)\frac{\q}{1-\q}\,\cdotp\ee
It is straightforward to check that \eqref{aqrel} is consistent with
\eqref{Feq} and thus equivalent to \eqref{rel} for $p=1$, $\mathscr
P_{1}=\mathscr P_{1}^{(0)}$.

\subsection{A useful lemma}
\label{lemmaSec}

To proceed further, we need a simple algebraic\smallskip\\
\textbf{Lemma}: \emph{Let $n$ be a positive integer. Let $|i\rangle$,
$i\in I$, a subset of vacua, with classical limits} $|i\ranglecl$.
\emph{Assume that we can prove, for all $A\in\mathcal V_{n}$, that}
$A\bigl(\langle i|\x|i\ranglecl,\b,\q\bigr) = 0$ \emph{for all $i$
implies that $A$ identically vanishes. Then if $P\in\mathcal V_{n}$ is
such that $P\bigl(\langle i|\x|i\rangle,\b,\q\bigr) = 0$ for all $i$,
$P$ must identically vanish.}\smallskip\\
This result is very useful, because the classical expectation values
$\langle i|\x|i\ranglecl$ are much simpler that their quantum
counterparts $\langle i|\x|i\rangle$.

To prove the lemma, we consider $P\in\mathcal V_{n}$ such that
\be\label{Lem1} P\bigl(\langle i|\x|i\rangle,\b,\q\bigr) = 0\ee
for all $i\in I$. We expand
\be\label{Lem2} P(\x,\b,\q) = \sum_{k\geq 0}A_{k}(\x,\b)\,\q^{k}\, ,\ee
where $A_{k}\in\mathcal V_{n}$. Equation \eqref{Lem1} is equivalent to
\be\label{Lem3} \sum_{k\geq 0}A_{k}\bigl(\langle
i|\x|i\rangle,\b\bigr)\,\q^{k} = 0\, .\ee
Note that, of course, $\langle i|\x|i\rangle$ depends on $\q$ in
general. 

Let us show recursively on $k$ that \eqref{Lem3} implies
\be\label{Lem4} A_{k} = 0\ee
for all $k\geq 0$. The vanishing of $P$ will follow immediately. To
prove the case $k=0$, let us take the $\q\rightarrow 0$ limit of
\eqref{Lem3},
\be\label{Lem3b} A_{0}\bigl(\langle i|\x|i\ranglecl,\b\bigr) = 0\,
.\ee
The vanishing of $A_{0}$ then follows from the basic assumption in the
lemma. Assume now that \eqref{Lem4} is valid for $k\leq k_{0}$.
Equation \eqref{Lem3} then yields
\be\label{Lem5} \sum_{k\geq k_{0}+1}A_{k}\bigl(\langle
i|\x|i\rangle,\b\bigr)\,\q^{k} = 0\, ,\ee
which implies that
\be\label{Lem6}A_{k_{0}+1}\bigl(\langle i|\x|i\rangle,\b\bigr) +
\sum_{k\geq 1}A_{k_{0}+1+k}\bigl(\langle
i|\x|i\rangle,\b\bigr)\,\q^{k} = 0\, .\ee
We deduce that \eqref{Lem4} is valid for $k=k_{0}+1$ by taking the
$\q\rightarrow 0$ limit and applying again the basic assumption in the
lemma.

\subsection{Using the rank one vacua}
\label{rankoneSec}

For the vacua of rank one, the curve \eqref{Cdef} is a sphere and
$\langle F(z)\rangle_{r=1}$, which is obtained from $\langle
R(z)\rangle_{r=1}$ by performing elementary integrals, automatically
satisfies a degree two algebraic equation. The solution is
parametrized by a single unknown parameter, the glueball expectation
value $\langle S\rangle_{r=1}=\langle v_{0}\rangle_{r=1}/N$. It is
completely elementary to check that
\be\label{Fr1eq} \langle F(z)\rangle_{r=1} + \frac{h U(z)}{\langle
F(z)\rangle_{r=1}} = \langle H(z)\rangle_{r=1}\ee
for some polynomial $H$ and some function $h$ of $\langle
S\rangle_{r=1}$ and of the parameters. For example, if we use a
quadradic superpotential $W$, i.e.\ $g_{k}=0$ for $k\geq 2$ (this is
not a restriction because the relations \eqref{defPquantum} do not
depend on the $g_{k}$s), we find that
\be\label{hquadfor} h =
\frac{2^{2N-2\sum_{Q}\nu_{Q}}}{g_{1}^{N-\sum_{Q}\nu_{Q}}U_{0}}\langle
S\rangle^{N-\sum_{Q}\nu_{Q}}\prod_{Q}
\Bigl(b_{Q}+\sqrt{b_{Q}^{2}-4\langle S\rangle/g_{1}}\Bigr)^{2\nu_{Q}-1}\,
.\ee
The $\nu_{Q}$s were defined in \ref{modelSec}.

We can now use the operator relation \eqref{uN1rel}, that we have
derived using the rank zero vacua, to fix the function $h$. We obtain
that $U_{0}h$ must be a function of $\q$ only,
\be\label{ahrel} U_{0}h\bigl(\langle S\rangle_{r=1},\g,\b\bigr) =
\frac{\mathsf a(\q)}{N+1+\mathsf a(\q)}\Longleftrightarrow \mathsf
a(\q) = (N+1)\frac{U_{0}h}{1-U_{0}h}\,\cdotp\ee
Even though we do not really need it, let us briefly explain how the
precise relation between $\mathsf a$ and $\q$, equation \eqref{aqrel},
can be obtained. The idea is to compute the glueball superpotential.
On the one hand, as reminded in \ref{crctSec}, this superpotential is
fixed by the anomaly equations modulo the addition of an arbitrary
function $f(S)$ that depends on the glueball field $S$ but not on the
couplings $\g$, $\b$ or $\q$. On the other hand, the glueball
superpotential can be computed unambiguously from $\langle
S\rangle_{r=1}$ which is given by \eqref{ahrel}. It is then
straightforward to check that consistency between the two results
implies \eqref{aqrel}. A very simple way to understand why this must
be valid, without performing any explicit calculation, is as follows
\cite{ferproof}. The equation \eqref{ahrel} has been obtained by
implementing consistently the constraints from the $\u_{\text{R}}$
symmetry of the theory, see the charge asignments \eqref{asign}. This
symmetry also implies that the glueball superpotential must satisfy
the differential equation
\be\label{uRsym} S\frac{\partial W}{\partial S} + \sum_{k\geq
0}g_{k}\frac{\partial W}{\partial g_{k}} = W\, .\ee
As emphasized in the Section 4 of \cite{ferproof}, this differential
equation fixes the coupling-independent part $f(S)$ in $W(S,\g,\b,\q)$
up to a linear term in $S$ that corresponds to an overall numerical
factor that may multiply $\q$. In this approach, the numerical factor
can be fixed by performing a single one-instanton calculation, for
example in the Coulomb vacuum discussed in \ref{rankNSec}, and one
finds again that $h=q$.

So we know that the relation \eqref{Feq} is valid in the rank one
vacua. Using the decomposition \eqref{reldec}, this is equivalent to
the constraints
\be\label{P1vanish2}\mathscr
P_{p}^{(1)}\bigl(\langle\x\rangle_{r=1},\b,\q\bigr) = 0\, ,\quad p\geq
1\, ,\ee
which is similar to \eqref{P1vanish}, but now for the rank one vacua.
To analyse the algebraic consequences of \eqref{P1vanish2}, we shall
use the lemma of Section \ref{lemmaSec}. To do this, let us first
describe the classical limits of the rank one vacua. They correspond
to having $p$ of the $x_{i}$s, say $x_{1},\ldots,x_{p}$, to be equal
to $p$ distinct $b_{j}$s, for example $x_{i}=b_{i}$ for $1\leq i\leq
p$, and to having all the other $x_{i}$s, $i>p$, to be equal to the
same root $w$ of the polynomial $W'$ given in \eqref{wtdef},
$x_{p+1}=\cdots=x_{N}=w$. The roots of $W'$ are algebraically
independent from $\b$ and $\q$, and thus can be considered to be
arbitrary indeterminates for our purposes. We are now going to prove a
result which is the analogue, for the rank one vacua, of the
proposition of Section \ref{rank0Sec}:\smallskip\\
\textbf{Proposition}: \emph{Let $A\in\mathcal V_{n}$, $A\not = 0$,
such that}
\be\label{ConstA2}
A(x_{1}=b_{1},\ldots,x_{p}=b_{p},x_{p+1}=w,\ldots,x_{N}=w,\b,\q)
= 0\ee
\emph{for all $0\leq p\leq N-1$. Then $\deg A=n\geq
2N+6$.}\smallskip\\
The proof is very similar to the one given in \ref{rank0Sec} after
\eqref{AN1}. The assumptions in the proposition imply that there
exists $k_{0}\leq N-1$, defined to be the smallest integer such that
$A(x_{1}=b_{1},\ldots,x_{k_{0}}=b_{k_{0}},
x_{k_{0}+1},\ldots,x_{N},\b,\q)$ vanishes. Seeing
$A(x_{1}=b_{1},\ldots,x_{k_{0}-1}=b_{k_{0}-1},
x_{k_{0}},\ldots,x_{N},\b,\q)$ as a polynomial in $x_{k_{0}}$, using
the symmetry in the $b_{k}$s for $k\geq k_{0}$ and then using the
symmetry in the $x_{i}$s for $k_{0}\leq i\leq N$, we deduce that
\be\label{Adedr2} A(x_{1}=b_{1},\ldots,x_{k_{0}-1}=b_{k_{0}-1},
x_{k_{0}},\ldots,x_{N},\b,\q) =
\prod_{i=k_{0}}^{N}\prod_{k=k_{0}}^{2N}(x_{i}-b_{k}) B\ee
for some non-zero $B$. Using the homogeneity of $A$, we get
\be\label{const2degA} \deg A\geq (2N-k_{0}+1)(N-k_{0}+1)\, .\ee
This yields
\be\label{cdegA3} \deg A\geq 3(N+3)\quad\text{if}\ k_{0}\leq N-2\,
.\ee
If $k_{0}=N-1$, \eqref{ConstA2} implies that
$A(x_{1}=b_{1},\ldots,x_{N-2}=b_{N-2},x_{N-1},x_{N})$ not only
vanishes at $x_{N-1}=b_{i}$ for $i\geq N-1$, but also at
$x_{N-1}=x_{N}$. Using in particular the symmetry in exchanging
$x_{N-1}$ and $x_{N}$, we get
\begin{multline}\label{Adedr2bis}A(x_{1}=b_{1},\ldots,
x_{N-2}=b_{N-2},x_{N-1},x_{N}) =\\
(x_{N-1}-x_{N})^{2}\prod_{k=N-1}^{2N}(x_{N-1}-b_{k})(x_{N}-b_{k})
B\end{multline}
for some non-zero $B$, and thus
\be\label{cdegA4} \deg A\geq 2(N+2)+2=2N+6\quad\text{if}\ k_{0}=N-1\,
.\ee
Together with \eqref{cdegA3}, this proves the proposition.

An immediate corrolary is that, if $A\in\mathcal V_{n}$ satisfies
\eqref{ConstA2} and $n\leq 2N+5$, then $A=0$. We can thus apply the
lemma of Section \ref{lemmaSec}, for the subset $I$ of all the rank
two vacua, to deduce from \eqref{P1vanish2} that $\mathscr
P_{p}^{(1)}$ must vanish if $\deg\mathscr P_{p}^{(1)}= N+p\leq 2N+5$,
or $p\leq N+5$.

Let us summarize what we have done. The anomaly equations imply that
the relations
\be\label{relvac1} \langle u_{N+p}\rangle_{r=1} = \mathscr
P_{p}^{(0)}\bigl(\langle u_{1}\rangle_{r=1},\ldots,\langle
u_{N}\rangle_{r=1},\b,\q) \ee
are valid for all $p\geq 1$ in the rank one vacua. This implies that
the relations
\be\label{relop} u_{N+p} = \mathscr P_{p}^{(0)}\bigl( u_{1},\ldots,
u_{N},\b,\q) \ee
are valid as \emph{operator relations} for
\be\label{pint1} 1\leq p\leq N+5\, .\ee
Equivalently, \eqref{Feq} must be valid as an operator relation up to
terms of order $z^{-N-6}$,
\be\label{relopbis} F(z) + \frac{qU(z)}{F(z)} = H(z) + \mathcal
O\bigl(1/z^{N+6}\bigr)\ee
for some degree $N$ polynomial $H$.

\subsection{Using the rank two vacua}

Operator relations are valid in all the vacua, and thus
\eqref{relopbis} yields $N+5$ non-trivial constraints on the solutions
\eqref{gensolR}. This is enough the fix completely $\langle
R(z)\rangle_{r}$ as long as the number of parameters is smaller than
the number of constraints, i.e.\ when $2r-1\leq N+5$. In particular,
we know all the rank two solutions.

Concretely, taking the derivative of \eqref{relopbis} with respect to 
$z$, we find that \eqref{Rform} must be valid in the rank two vacua 
up to terms of order $1/z^{2N+7}$. Comparing with \eqref{gensolR}, we 
thus obtain
\begin{multline}\label{r2const} \frac{C}{y_{2}} +
\frac{1}{2}\frac{U'}{U} -
\frac{1}{2y_{2}}\sum_{Q}\frac{(1-2\nu_{Q})y_{2}(z=b_{Q})}{z-b_{Q}} =\\
\frac{1}{2}\frac{U'}{U} + \Bigl(H'_{2} -
\frac{U'H_{2}}{2U}\Bigr)\frac{1}{\sqrt{H_{2}^{2} - 4qU}} +\mathcal
O\bigl(1/z^{2N+7}\bigr)\, ,
\end{multline}
for some constant $C$ and where we have defined $H_{2} = \langle
H\rangle_{2} = (1+\q)z^{N}+\cdots$. By expanding at large $z$,
\eqref{r2const} yields $2N+5$ non-trivial constraints, which is more
than enough to determine the $N+3$ free parameters in $y_{2}$, $C$ and
$H_{2}$ (actually, we only need \eqref{r2const} up to terms of order
$1/z^{N+5}$). We have checked explicitly that the solution is indeed
uniquely fixed, in the particular case of the rank two vacuum
corresponding to the classical limit
\be\label{R2cl}\bigl\langle R(z)\bigr\rangle_{2,\,\text{cl}} =
\sum_{i=1}^{N-2}\frac{1}{z-b_{i}} +
\frac{1}{z-w_{1}}+\frac{1}{z-w_{2}}\,\cdotp \ee
Note that we shall use only this vacuum in the following. Performing
the check is completely straightforward, but quite tedious. The idea
is to study $\langle R(z)\rangle_{2}$ in a small $q$ expansion around
the classical solution \eqref{R2cl}, using \eqref{r2const}. A
recursive argument shows that the expansion parameter is $q$ (not a
fractional power of $q$), as expected in this weakly coupled vacuum
with unbroken gauge group $\u^{2}$, and that the coefficients in the
small $q$ expansion are uniquely fixed to all orders by the
constraints \eqref{r2const}. We have not tried, however, to work out
directly the explicit form of the solution to all orders from
\eqref{r2const}. Actually, this is not necessary. A solution to
\eqref{r2const} is known, and corresponds to imposing the
factorization condition \eqref{fact2} at $r=2$. The uniqueness of the
solution then ensures that the solution obtained in the small $q$
expansion must correspond to this factorization condition. But the
factorization condition is equivalent to the validity of the
quantization conditions \eqref{QC1} and \eqref{QC2}, or to the fact
that \eqref{r2const} is actually true to \emph{all} orders, or also to
the relation
\be\label{Fr2eq} \langle F(z)\rangle_{2} + \frac{q U(z)}{\langle
F(z)\rangle_{2}} = \langle H(z)\rangle_{2}\ee
in the rank two vacua. Strictly speaking, we have proven \eqref{Fr2eq}
only in the rank two vacua with classical limits \eqref{R2cl}, but
this is enough for our purposes.

Using \eqref{reldec}, the equation \eqref{Fr2eq} is equivalent to
\be\label{P1vanish3}\mathscr
P_{p}^{(1)}\bigl(\langle\x\rangle_{2},\b,\q\bigr) = 0\, ,\quad p\geq
1\, .\ee
This is the rank two version of \eqref{P1vanish} and
\eqref{P1vanish2}. We can thus proceed along the lines of Sections
\ref{rank0Sec} and \ref{rankoneSec}. We shall use the: \smallskip\\
\textbf{Proposition}: \emph{Let $A\in\mathcal V_{n}$, $A\not = 0$,
such that}
\be\label{ConstA3}
A(x_{1}=b_{1},\ldots,x_{N-2}=b_{N-2},x_{N-1},x_{N},\b,\q)
= 0\, .\ee
\emph{Then $\deg A=n\geq 3N+9$.}\smallskip\\
The proof goes as after \eqref{AN1} or \eqref{ConstA2}. Using the
symmetry properties of the variables in $A\in\mathcal V_{n}$, the
constraints \eqref{ConstA3} imply that there exists $k_{0}\leq N-2$
such that
\be\label{Adedr3} A(x_{1}=b_{1},\ldots,x_{k_{0}-1}=b_{k_{0}-1},
x_{k_{0}},\ldots,x_{N},\b,\q) =
\prod_{i=k_{0}}^{N}\prod_{k=k_{0}}^{2N}(x_{i}-b_{k}) B\ee
for some non-zero $B$, and thus
\be\label{degreea} \deg A \geq (2N-k_{0}+1)(N-k_{0}+1)\geq 3(N+3)\ee
as we wished to show.

The condition \eqref{ConstA3} corresponds to the classical vacua
\eqref{R2cl} (note that the roots $w_{1}$ and $w_{2}$ of $W'$ are
algebraically independent from the $b_{Q}$s and $\q$, and thus play
the r\^ole of independent variables). We can thus use the lemma of
Section \ref{lemmaSec} to conclude that \eqref{P1vanish3} implies that
$\mathscr P_{p}^{(1)} = 0$ for all $p\leq 2N+8$, or equivalently that
\be\label{relopter} F(z) + \frac{qU(z)}{F(z)} = H(z) + \mathcal
O\bigl(1/z^{2N+9}\bigr)\ee
must be valid as an operator relation. The $2N+8$ non-trivial operator
relations that follow from \eqref{relopter} are more than enough to
fix unambiguously the free parameters in \eqref{gensolR}, in
\emph{all} the possible cases. Indeed, the maximal rank is $N$, and
the maximum number of parameters that can appear in \eqref{gensolR} is
thus $2N-1$.

\subsection{Using the rank N vacuum}
\label{rankNSec}

The proof of our main theorem is now at hand. Let us analyse the rank
$N$ Coulomb vacuum. Classically, this vacuum corresponds to
\be\label{Coulombcl} \bigl\langle R(z)\bigr\rangle_{N,\,\text{cl}} =
\sum_{i=1}^{N}\frac{1}{z-w_{i}}\, \cvp\ee
where the $w_{i}$s are the root of $W'$, see \eqref{wtdef}. Quantum
mechanically, the solution is uniquely fixed by \eqref{relopter} (the
validity of this equation is actually needed only up to terms of order
$1/z^{2N}$). This is shown as in the rank two case, using the analogue
of \eqref{r2const}. The unique solution must correspond to the known
one, which is characterized by the condition \eqref{fact2} at $r=N$
(in this case $\psi_{0}$ is just a constant and $y_{N}^{2} = W'^{2} -
4\Delta_{N-1}$). We deduce that the relation
\be\label{Fr3eq} \langle F(z)\rangle_{N} + \frac{q U(z)}{\langle
F(z)\rangle_{N}} = \langle H(z)\rangle_{N}\ee
is valid in the rank $N$ vacuum, or equivalently that
\be\label{P1vanish4}\mathscr
P_{p}^{(1)}\bigl(\langle\x\rangle_{N},\b,\q\bigr) = 0\, ,\quad p\geq
1\, .\ee
However, in the Coulomb vacuum, \emph{the} $\langle x_{i}\rangle$
\emph{are algebraically independent} from the $b_{Q}$s and $\q$ in the
classical limit \eqref{Coulombcl}. Combining this fact together with
\eqref{P1vanish4} and the lemma in Section \ref{lemmaSec}, we get
\be\label{final} \mathscr P_{p}^{(1)}\bigl(\x,\b,\q\bigr) = 0\, .\ee
This completes the proof of the chiral ring consistency theorem.

\section{Conclusions}
\setcounter{equation}{0}

The chiral ring consistency theorem sheds considerable light on the
inner workings of the gauge theory/matrix model correspondence. In the
matrix model, the planar limit must be taken and thus the variables
that enter the loop equations are all independent. As a consequence,
the most general solution is parametrized by arbitrary filling
fractions. In the gauge theory, the number of colours $N$ is finite,
relations like \eqref{defPquantum} must exist, and there is only a
finite number of independent variables. Consistency between the matrix
model loop equations, that are mapped onto the gauge theory
generalized Konishi anomaly equations, and the gauge theory identities
\eqref{defPquantum}, is then possible only for some particular values
of the filling fractions. These correspond to the expectation values
of the gauge theory glueball superfields and encode a very rich
non-perturbative dynamics.

As explained in the introduction, our results also illustrate a deep
consistency property of the open/closed string duality. The closed
string results can be written in the open string language if and only
if the closed string superpotential is extremized. Algebraic
identities in the open string picture and closed string equations of
motion are exchanged in the duality.

The general line of thinking used in the present paper was already
used in \cite{ferchiral}. The argument in this earlier work was that
the relations \eqref{defPquantum} can be determined by looking at the
weakly coupled Coulomb vacuum, because of the algebraic independence
of the variables in this case. At least in principle, everything can
be computed in this vacuum by performing explicit instanton
calculation. Since the relations \eqref{defPquantum} are operator
equations, they must then be valid in all the other vacua of the
theory, including the strongly coupled vacua where the semi-classical
approximation does not apply. Since they are equivalent to the
quantization conditions \eqref{QC1} and \eqref{QC2}, the latter must
also be valid in all the vacua of the theory. The main contribution of
the present paper, with respect to \cite{ferchiral}, is to show that
the explicit calculations in the Coulomb vacuum, which require
considerable technology, are not necessary if one starts from the
non-perturbative anomaly theorem. Everything is then fixed by the
internal algebraic consistency of the chiral ring.

We believe that the general philosophy of the present work applies to
\emph{any} $\nn=1$ supersymmetric gauge theory, including in the cases
where there is a moduli space of vacua. By combining the quantum
version of the classical equations of motion written in terms of the
gauge invariant observables, which are the generalized Konishi anomaly
equations, with the full set of identities that follow from the
definition of these variables in terms of fields transforming
non-trivially under the gauge group, one should be able to determine
unambiguously all the quantum vacua and associated chiral operators
expectation values.

\subsection*{Acknowledgements}

This work is supported in part by the belgian Fonds de la Recherche
Fondamentale Collective (grant 2.4655.07), the belgian Institut
Interuniversitaire des Sciences Nucl\'eaires (grant 4.4505.86), the
Interuniversity Attraction Poles Programme (Belgian Science Policy)
and by the European Commission FP6 programme MRTN-CT-2004-005104 (in
association with V.\ U.\ Brussels). Vincent Wens is a junior
researcher (Aspirant) at the belgian Fonds National de la Recherche
Scientifique. Frank Ferrari is on leave of absence from the Centre
National de la Recherche Scientifique, Laboratoire de Physique
Th\'eorique de l'\'Ecole Normale Sup\'erieure, Paris, France.

\renewcommand{\thesection}{\Alph{section}}
\renewcommand{\thesubsection}{\arabic{subsection}}
\renewcommand{\theequation}{A.\arabic{equation}}
\setcounter{section}{0}
\end{document}